\documentclass[twocolumn,showpacs,amsmath,amssymb]{revtex4}

\usepackage{graphicx}
\usepackage{dcolumn}
\usepackage{bm}
\usepackage{epsf}
\usepackage{epsfig}
\usepackage{graphics}

\begin{document}

\title{Afshar's Experiment does not show a Violation of Complementarity}

\author{Ole Steuernagel\\
\email{ole@star.herts.ac.uk}
 {\em School of Physics, Astronomy and Mathematics,
University of Hertfordshire, 
Hatfield, AL10 9AB, UK}}

\date{\today}

\begin{abstract}
A recent experiment performed by S. Afshar [first reported by M.
Chown, New Scientist {\bf 183}, 30 (2004)] is analyzed. It was
claimed that this experiment could be interpreted as a
demonstration of a violation of the principle of complementarity
in quantum mechanics. Instead, it is shown here that it can be
understood in terms of classical wave optics and the standard
interpretation of quantum mechanics. Its performance is quantified
and it is concluded that the experiment is suboptimal in the sense
that it does not fully exhaust the limits imposed by quantum
mechanics.
\end{abstract}



\maketitle

%
\section{Introduction}
%
Bohr's principle of complementarity of quantum mechanics
characterizes the nature of a quantum system as being dualistic
and mutually exclusive in its particle and wave
aspects~\cite{Wheeler.buch}. In the famous debates between
Einstein and Bohr in the late 1920's~\cite{Wheeler.buch}
complementarity was contested but finally the argument has settled
in its favour~\cite{Wheeler.buch,Feynman.buch}. Fourty years later
Feynman stated in his 1963 lectures quite categorically that
``\emph{No one has ever found (or even thought of) a way around
the uncertainty principle}"~\cite{Feynman.Uncertainty}.

Some recent discussions centered on the question of whether the
principle of complementarity is founded on the uncertainty
principle~\cite{Scully91.nature,Zajonc91,Tan93,Storey94,Englert95.nature,Storey95.nature,Wiseman95.nature,Englert96}.
Irrespective of whether or not one subscribes to Feynman's point
of view that complementarity and uncertainty are essentially the
same thing~\cite{Bhandari92,Bjoerk99,myQphys}, there is agreement
that the principle of complementarity is at the core of quantum
mechanics.

A step towards the quantification of the principle can be found in
reference~\cite{Wooters79} which has more recently inspired a
formulation using visibility~$V$ of the interference pattern as
the quantification of the wave nature of a quantum particle and
the difference between path detector states as the
path-distinguishability measure~$\cal D$. Together they give rise
to an inequality
%
\begin{eqnarray}
V^2 + {\cal D}^2 \leqq 1 \; ,
\label{EnglertIneq}
\end{eqnarray}
%
derived by Englert in reference~\cite{Englert96}. Here, this
inequality will be taken as the basis for a quantitative
discussion of the principle of complementarity.

A recent laser experiment performed by S. Afshar has been
presented as a possible counterexample to the principle of
complementarity in quantum mechanics, in particular
inequality~(\ref{EnglertIneq}) was supposedly
violated~\cite{Chown.04,Afshar05}. Since some confusion has arisen
in the context of the interpretation of Afshar's
experiment~\cite{Chown.04,Afshar05,Kastner05,Drezet05,press.echo}
a classical wave-optical analysis is given to explain the
mechanism of the experiment and derive the scattering amplitudes
for all partial waves involved. Subsequently, these amplitudes are
used in a quantum calculation to demonstrate that a quantitative
description is straightforward and that, far from violating the
principle of complementarity, the Afshar experiment is less than
optimal in the sense that it does not fully exhaust the limits of
quantum mechanics prescribed by inequality~(\ref{EnglertIneq}).

Afshar's experimental setup and procedure are introduced in the
next section. This is followed by a wave optical analysis in
section~\ref{sec.3.wave.optical.analysis} which is subdivided into
an informal wave optical description of the experiment in
subsection~\ref{subs.3a.qualitative} and a quantitative analysis
of the scattering amplitudes in
subsection~\ref{subs.3b.quantitative} and their effects on the
observed intensities. After that a quantification of the principle
of complementarity is used in
section~\ref{sec.5.quantify.complementarity} to prove that the
Afshar experiment does not `violate quantum mechanics', this is
followed by the conclusion.
%
\section{Afshar's experiment and his interpretation \label{sec.2.Setup}}
%
\begin{figure}[h t]
\centering
\includegraphics[width=3.4in,height=1.5in]{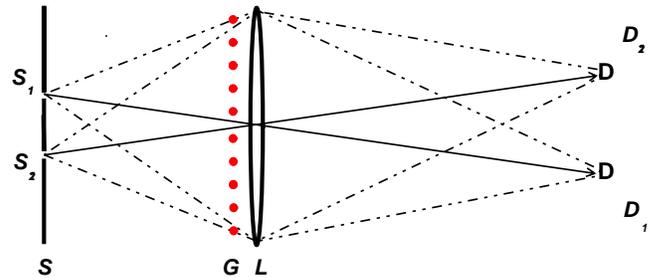}
\caption{Setup of the `Afshar experiment'
\cite{Chown.04,Afshar05}: a double-slit~$S$ is coherently and with
equal intensity illuminated from the left. With the help of
lens~$L$ the slits' images are recorded by detectors~$D_1$
and~$D_2$. At first, measurements are made \emph{without} grid~$G$
(dotted line); subsequently, the grid is inserted and the
measurements repeated. This sketch is a slightly simplified
version of the original which contains an aperture around lens~$L$
and redirection mirrors in front of the detectors. Both
simplifications are unimportant for the essentials of the
experiment.
\label{fig1}}
\end{figure}
%
\emph{Stages of the experiment:} in stage (i) of the experiment
the grid $G$ is not present. In this stage only one slit is left
open, therefore, only the corresponding detector gets illuminated.
Opening of the other slit (and illuminating the other detector
simultaneously) does not change the image at the first detector.

In stage (ii) the setup is modified by inserting the grid~$G$
directly in front of the lens~$L$, as shown in Fig.~\ref{fig1}.
The grid is carefully positioned in such a way that the wires sit
at the minima of the interference pattern (which can be checked by
inserting a screen at the position of $G$ and opening the second
slit). It is then observed that if one of the two slits is blocked
the image of the remaining open slit is strongly modified due to
the presence of the grid. The wires of the grid reflect (and
diffract) some of the photons, which leads to a reduction of the
intensity and introduces the formation of stripes in the image of
the slit~\cite{Chown.04,Afshar05}. Note that the positioning of
the grid at the minima of the interference pattern implies that it
constitutes a carefully matched \emph{diffraction grating} for the
passing light.

In stage (iii) the other slit (previously blocked) is reopened.
The crucial result at this stage is that the focal images of the
two slits at detectors $D_1$ and $D_2$ look remarkably similar to
those obtained in the absence of the grid in stage~(i).
\\

Afshar's interpretation, as cited from reference~\cite{Chown.04}:
\begin{itemize}
\item[]\emph{
Laser light falls on two pinholes in an opaque screen. On the far
side of the screen is a lens that takes the light coming through
each of the pinholes (another opaque screen stops all other light
hitting the lens) and refocuses the spreading beams onto a mirror
that reflects each onto a separate photon detector. In this way,
Afshar gets a record of the rate at which photons are coming
through each pinhole. According to complementarity, that means
there should be no evidence of an interference pattern. But there
is, Afshar says.}

\emph{He doesn't look at the pattern directly, but has designed
the experiment to test for its presence. He places a series of
wires exactly where the dark fringes of the interference pattern
ought to be. Then he closes one of the pinholes. This, of course,
prevents any interference pattern from forming, and the light
simply spreads out as it emerges from the single pinhole. A
portion of the light will hit the metal wires, which scatter it in
all directions, meaning less light will reach the photon detector
corresponding to that pinhole.}

\emph{But Afshar claims that when he opens up the closed pinhole,
the light intensity at each detector returns to its value before
the wires were set in place. Why? Because the wires sit in the
dark fringes of the interference pattern, no light hits them, and
so none of the photons are scattered. That shows the interference
pattern is there, says Afshar, which exposes the wave-like face of
light. And yet he can also measure the intensity of light from
each slit with a photon detector, so he can tell how many photons
pass through each slit - the particle-like face is there too.}

\emph{"This flies in the face of complementarity, which says that
knowledge of the interference pattern always destroys the
which-way information and vice versa," says Afshar. "Something
everyone believed and nobody questioned for 80 years appears to be
wrong."}
\end{itemize}

In other words, stage (iii) of the experiment is interpreted as a
demonstration of simultaneous `perfect' particle ($D=1$) and wave
behaviour ($V=1$) thus violating the complementarity
inequality~(\ref{EnglertIneq})~\cite{Chown.04,Afshar05}.
%
\section{Wave Optical Analysis \label{sec.3.wave.optical.analysis}}
Afshar's interpretation relies on the fact that the slits' images
obtained in the presence of the grating, in stage~(iii), are very
similar to those obtained in stage (i) without a grating. In the
article~\cite{Chown.04} they are describes in the figure caption
of the slit images in stage~(iii) of the experiment as returning
to their `original value', that of stage~(i). But, close
inspection of the images themselves reveals that in stage~(iii)
they display residues of the same disturbances that are seen in
stage~(ii)~\cite{Chown.04,Afshar05}. These disturbances were
considered negligible and without fundamental significance. Here,
it is shown that these residual disturbances must not be
neglected, they are key to understanding and interpreting Afshar's
experiment.
\subsection{Qualitative analysis \label{subs.3a.qualitative}}
Putting the grid into the area where the dark fringes are to be
expected in stage~(iii) of Afshar's experiment is an elegant way
of proving that some interference contrast is present, while quite
efficiently avoiding the reflection or absorption of passing
photons by the grid. But avoiding the reflection or absorption of
photons is not enough to guarantee that their behaviour in
relation to the complementarity principle is unaffected. The
grid's other main effect is the elastic scattering of passing (and
reflected) photons due to \emph{diffraction}. Even if the grid $G$
was perfectly absorbing, and hence did not reflect any photons,
would it still \emph{diffract passing photons} on their way from
slit $S_1$ to detector $D_1$. Indeed, quite a few of those photons
do get scattered towards detector~$D_2$ because it lies in the
direction of the first diffraction order of the grating, see
below. For symmetry reasons, analogous perturbations affect
photons on their way from slit $S_2$ to $D_2$, redirecting some
towards detector~$D_1$.

This explains some key features of Afshar's experiment. In
stage~(ii) of the experiment reflection and absorption, and
diffraction by the grid distort the slit images at the detectors.
In stage~(iii) the other slit is opened and first-order
diffraction of photons from that newly opened slit apparently
restores the slit images. It also shows qualitatively how the
complementarity principle is at work: the path detection in the
presence of the grid becomes less reliable since photons are
diffracted towards the `wrong' detector thus compromising path
detection.
\\
%
\begin{figure}[h b]
\centering
\includegraphics[width=3.2in,height=1.6in]{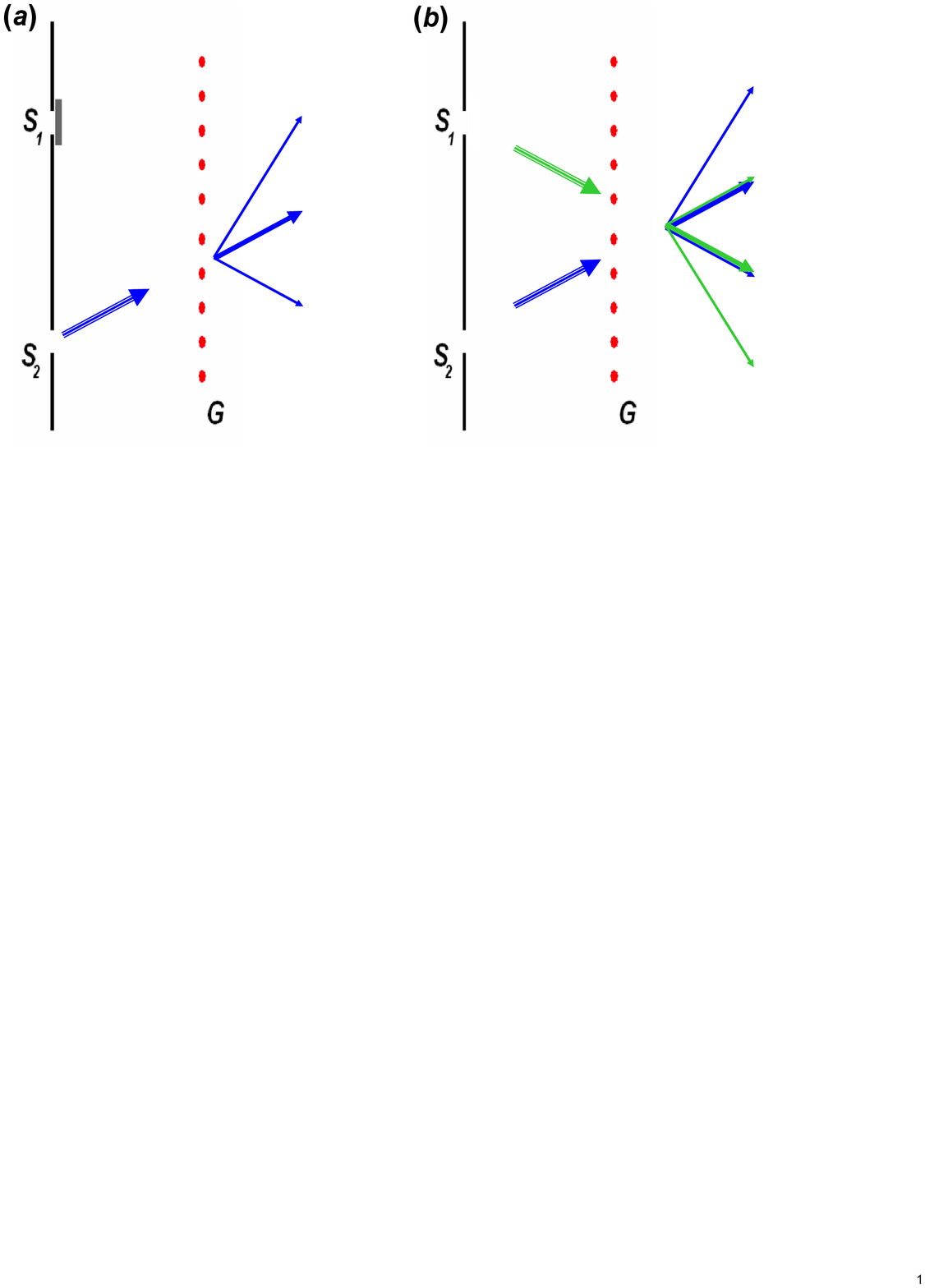}
\caption{Geometry of wave vectors in the vicinity of the grid
plane~$G$: \textbf{(a)} only slit $S_2$ is open, \textbf{(b)} both
slits are open. Some forward scattered light is deflected sideways
by the grid~\textbf{(a)}. This loss of forward scattered light is
partly compensated for by opening the other slit~\textbf{(b})
which `adds on' more light with the correct phase and transverse
momentum. Here, only zeroth and first order transmitted light wave
vectors are plotted.
\label{_2_wave_vectors}}
\end{figure}
\subsection{Quantitative analysis of diffraction\label{subs.3b.quantitative}}

In the forthcoming analysis we introduce three simplifications
that do not affect the underlying mechanism of the experiment.
Firstly, we assume the light is monochromatic with angular
frequency $\omega = c \, k$ and wave number $k$, and we only
analyze features in the plane defined by the optical axis and the
two slits $S_1$ and $S_2$, that is, we treat the problem in two
spatial dimensions. Secondly, we assume that the we can use the
paraxial wave approximation, namely, we describe the light in
terms of plane waves and all angles are small. Later, it will be
confirmed that these two simplifications are helpful without
affecting the generality of the argument. Thirdly, we model the
grid by a planar reflecting film consisting of suitable strips of
metal rather than an array of reflecting round wires. This last
assumption would lead to a slightly better performance than
Afshar's setup, using round wires, because reflection (although
not diffraction) of photons towards the `wrong' detector would
become completely suppressed. It also simplifies our analysis
because we now encounter only two mutually exclusive subensembles
of photons, transmitted and back-reflected ones.

\begin{figure}[h t]
\centering
\includegraphics[width=1.6in,height=3.1in,angle=-90]{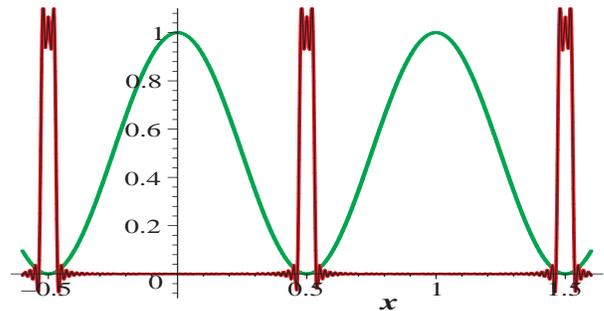}
\caption{The grid function~$G(x)$ of eq.~(\ref{grating}) for a
grid with covering ration $a=0.06$ described by a Fourier series
expanded up to 50th order together with the interference pattern
intensity $\cos(k_{\perp} x)^2$ of the two slits $S_1$ and $S_2$
in the grid plane. The $x$-axis is scaled in terms of the grid
spacing~$\Lambda$.
\label{plot_grid}}
\end{figure}

With these simplifying assumptions we can determine the
interference pattern at the grid to be proportional to
$\cos(k_{\perp} x)^2 = \cos(\frac{k s}{g} x)^2$, where $k$ is the
wave number of the light and its transverse component $k_{\perp}=k
s / g $ arises from the geometry of the setup. The distance
between the slits is $2s$ and $g$ is the distance from the
double-slit to the grid; small angles (paraxial beams) are assumed
throughout. Consequently, the grid has spacing~$\Lambda= 2\pi / (2
k_\perp) = \pi g/(k s)$ and can be expanded into a Fourier-series
with periodicity $\Lambda$. Our model for the grid $G(x)$ is a
periodic comb of reflecting stripes; in other words, the
reflectivity alternates between values of unity in regions
centered at odd multiples of $\Lambda/2$ over a distance $\Lambda
a, (a<1)$ and zero over the remaining distance $\Lambda (1-a)$:
$a$ is the covering ratio of the grating. $G(x)$ is plotted,
together with the interference pattern, in Fig.~\ref{plot_grid}.
%
%

The grid's functional description in terms of a Fourier series is
given by
%
\begin{eqnarray}
G(x) & = & c_0 + \sum_{n=1}^{\infty} c_n \cos\left(\frac{2
\pi}{\Lambda} x n \right) \label{grating} \; ,\\
\mbox{with \quad} c_0 & = & a,\mbox{\quad and \quad}  c_n =
2(-1)^n\frac{\sin(a \pi n)}{\pi n}. \label{grating.coeffs}
\end{eqnarray}
%
We will now work out how this grating affects incoming plane
waves. For simplicity we assume that the wave vector of the
incident light~$\psi_{in}$ is perpendicular to the grating. We
will later include the slight deviation due to the fact that the
slits $S_1$ and $S_2$ lie half a diffraction order off the optical
axis. A plane wave travelling in the $z$-direction with wave
vector $\vec{k} = k\; {\bf{\hat{z}}}$ is described by $\psi_{in} =
e^{i(kz-\omega t)}$. When the grid diffracts this plane wave into
the $n$-th positive or negative order its wave vector becomes
$\vec{k}_{\pm n} = \pm k_{x,n}\; {\bf{\hat{x}}} + k_{z,n}\;
{\bf{\hat{z}}}$ where $k_{x,n}= n \cdot 2 k_\perp$
and $k_{z,n} = \sqrt{k^2 - k_{x,n}^2}$.

After interaction with the grating $G$ we thus find the
transmission light modes $\psi_{\pm n,t}=e^{i( k_{z,n} z \pm
k_{x,n} x -\omega t)}$, and the reflection modes $\psi_{\pm
n,r}=e^{i(-k_{z,n} z \pm k_{x,n} x -\omega t)}$. With the tacit
understanding that the grating function~(\ref{grating}) can be
viewed as an operator $\hat G$ that imparts transverse momentum
kicks of size $\hbar k_{x,n} = n \cdot \hbar 2 k_\perp \doteq n
p_\perp$, we thus introduce the corresponding momentum transfer
operator $\hat p_\perp$ which allows us to write down the effect
of the grating on an incoming wave as
%
\begin{eqnarray}
\hat G(x) & = & c_0 \hat 1  + \sum_{n=1}^{\infty} c_n \frac{e^{ i
\frac{\hat{ p_\perp}}{\hbar} x n}+e^{ -i \frac{\hat{
p_\perp}}{\hbar} x n}}{2}
\\
& = & a \hat 1  + \sum_{n=1}^{\infty}  \frac{(-1)^n \sin(a \pi
n)}{\pi n} \cdot \left[ e^{ i \frac{\hat{ p_\perp}}{\hbar} x
n}+e^{ -i \frac{\hat{ p_\perp}}{\hbar} x n} \right] . \quad
\label{gratingOperator}
\end{eqnarray}
%
Including the $\pi$ phase jump associated with
reflection~\cite{Feynman.Reflection} we therefore find for the
reflection amplitudes
%
\begin{eqnarray}
& & r_0=-c_0=-a \\
 \mbox{and} & & r_n=r_{-n}=\frac{-c_n}{2} = (-1)^{n+1} \frac{\sin(a \pi n)}{\pi n} \; .
\label{r0rn}
\end{eqnarray}
%
Following the same logic~\cite{Feynman.Reflection}, we find the
transmission amplitudes obey
%
\begin{eqnarray}
t_0=1+r_0 \; \mbox{  and  } \; t_n=t_{-n}=r_n \; .
\label{t0tn}
\end{eqnarray}
%
After a  multiplication with $e^{i \omega t}$ to remove the
time-dependence we thus arrive at the result that the reflected
and transmitted partial waves have the form
%
\begin{eqnarray}
[ \psi_t ] & + & [\psi_r]  =  \left[ (1-c_0)  e^{ikz} -
\sum_{n=1}^{\infty} \frac{c_n}{2} \cdot
\left( \psi_{+n,t} + \psi_{-n,t}\right) \right] \nonumber \\
& - & \left[ c_0  e^{-ikz}  + \sum_{n=1}^{\infty} \frac{c_n}{2}
\cdot \left( \psi_{+n,r} + \psi_{-n,r}\right) \right] \; . \quad
\;
\label{wave_Map}
\end{eqnarray}
%
With $\sum_{n=1}^{\infty} c_n^2 = 2(a - a^2)$ we can check the
normalization and find $\sum_{n=-\infty}^{\infty} (r_n^2 + t_n^2)
= r_0^2 + t_0^2 + \sum_{n=1}^{\infty} c_n^2 = a^2+(1-a)^2+
2(a-a^2)=1$, as required.

After the discussion of the effect of the grating on a single
plane wave $\psi_{in} = e^{i(kz-\omega t)}$ we can extend the
discussion to the case of two interfering plane waves as sketched
in Fig.~\ref{_2_wave_vectors} above. The transverse components
$\pm \hbar k_\perp = \pm \Delta p_\perp / 2$ of the incoming waves
sketched in Fig.~\ref{_2_wave_vectors}~{\bf (b)} offset the waves
by half orders up or down. This is why the diffracted (and
reflected) waves fall into half odd integer orders, compare e.g.
Fig.~\ref{figure5}~{\bf (a)}. The partial waves of some order~$n$
originating from one slit therefore overlap with the adjacent
order~$n-1$ or~$n+1$ partial waves from the other slit.

\begin{figure}[h t]
\centering
\includegraphics[width=1.0in,height=1.5in,angle=-90]{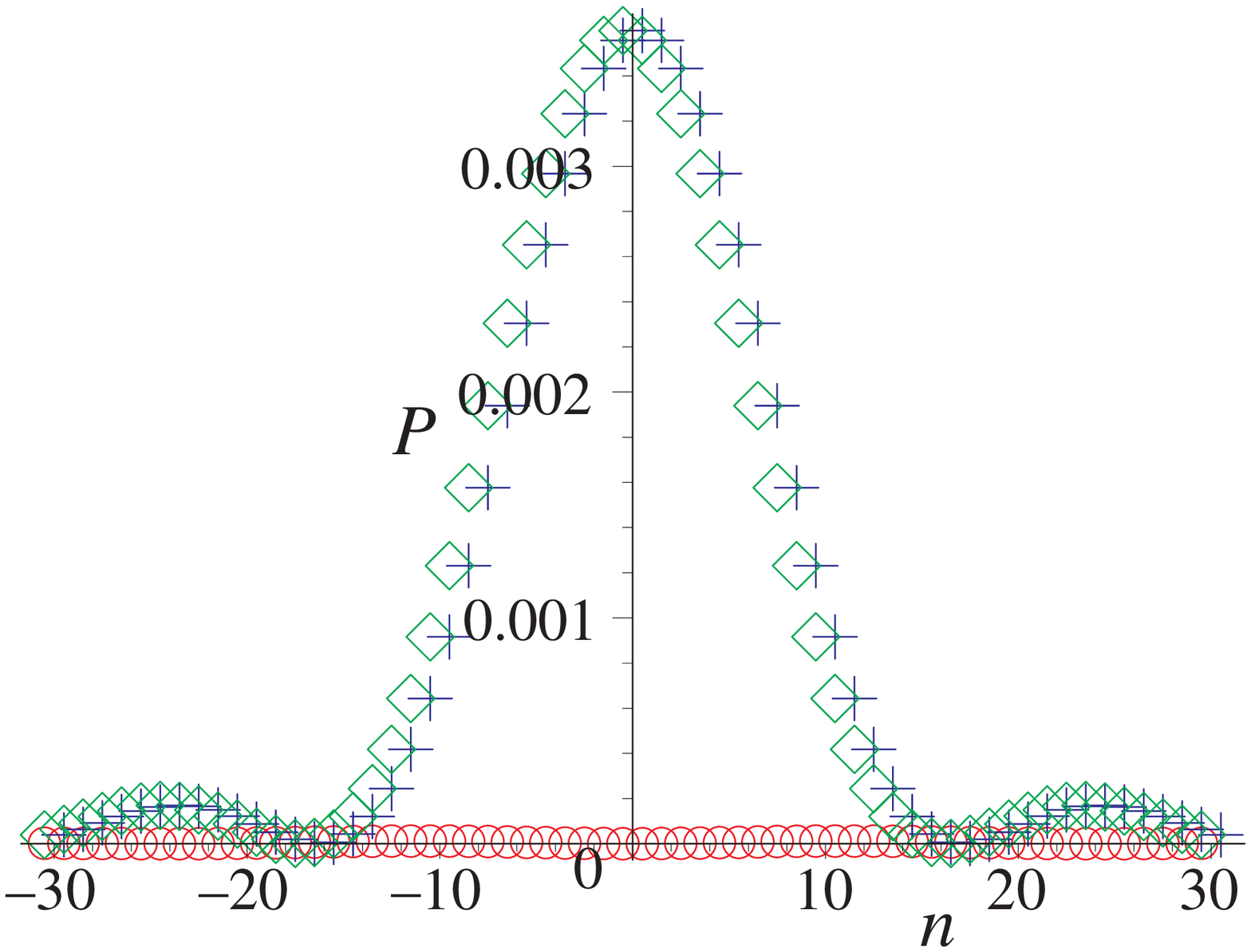}
\hspace{0.5cm}
\includegraphics[width=1.0in,height=1.5in,angle=-90]{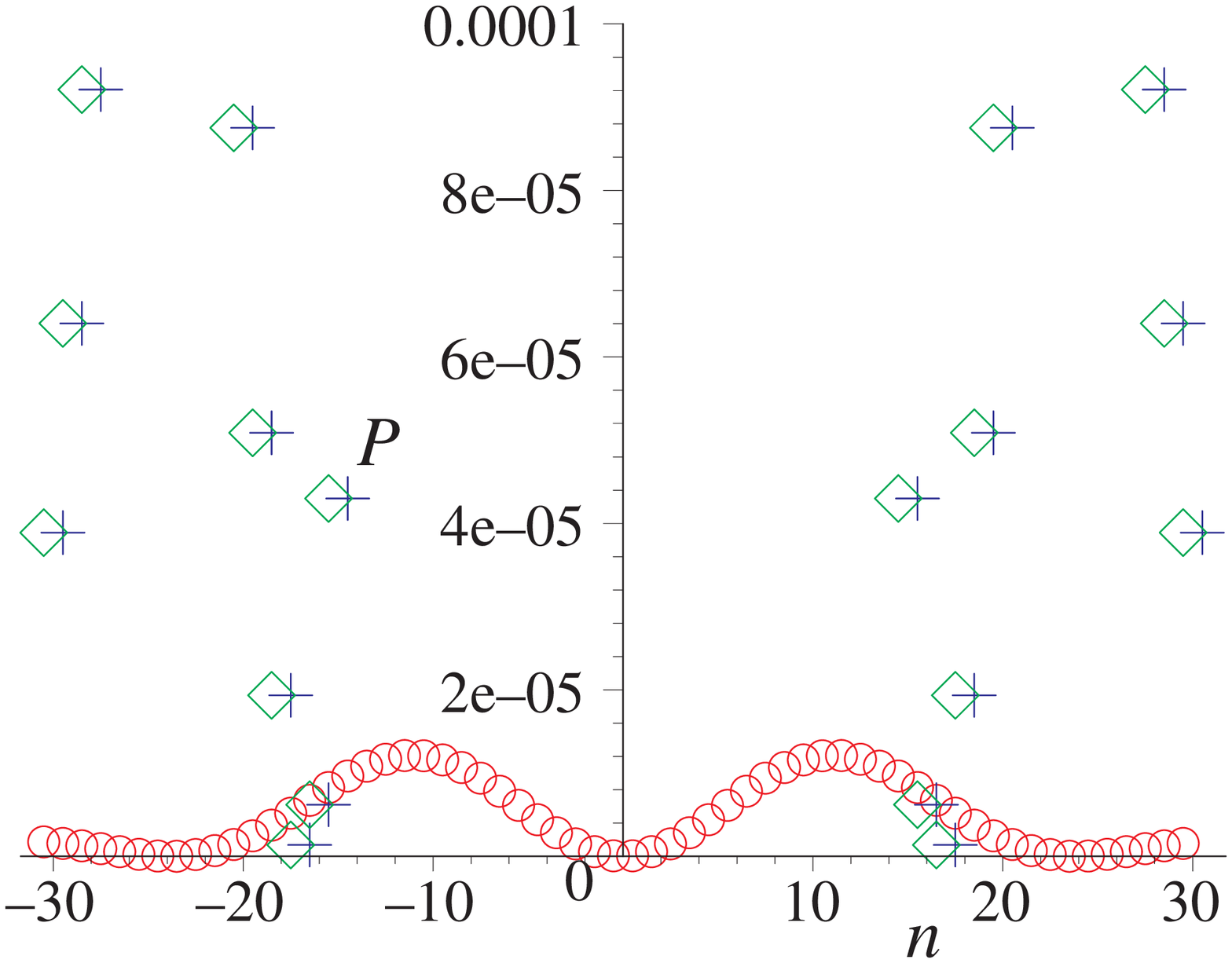}
\put(-220,-5){\small \bf (a)} \put(-90,-5){\small \bf (b)}
\caption{Reflection probabilities~$P=|r_n^2|$ up to 30th order for
a grid with covering ration $a=0.06$. Green diamonds and blue
crosses describe the reflection probabilities for individual beams
originating from slit $S_1$ and $S_2$ respectively, this
corresponds to stage~(ii) of the experiment, {\bf (a)}. When both
slits are opened simultaneously the reflection is very much
suppressed. One has to zoom in, {\bf (b)}, to see that the
reflection probability $P=|r_{n}+r_{n+1}|^2$ (red circles) in this
case is nonzero, this corresponds to stage~(iii) of Afshar's
experiment.
\label{figure4}}
\end{figure}

\begin{figure}[h t]
\centering
\includegraphics[width=1.2in,height=1.5in,angle=-90]{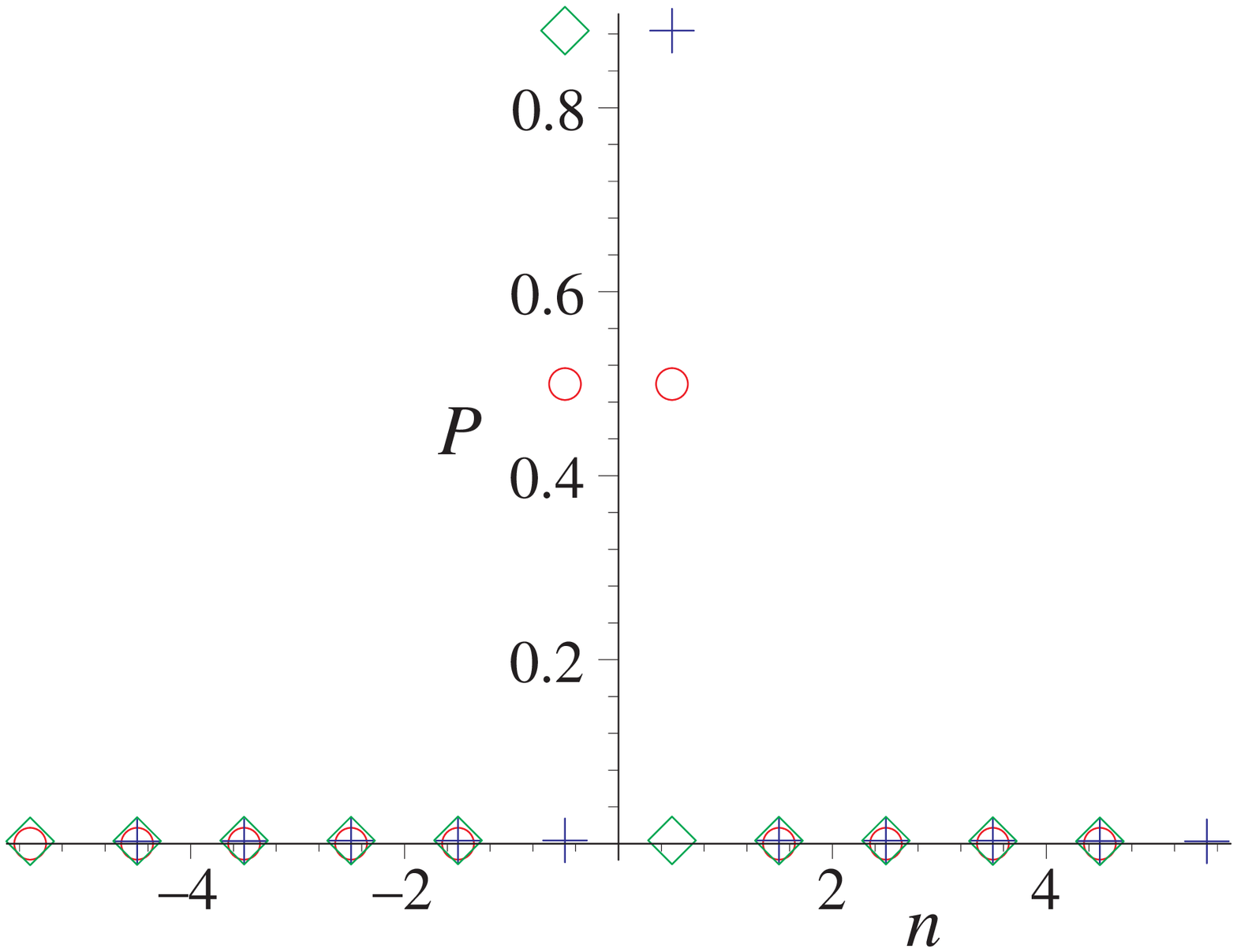}
\hspace{0.5cm}
\includegraphics[width=1.2in,height=1.5in,angle=-90]{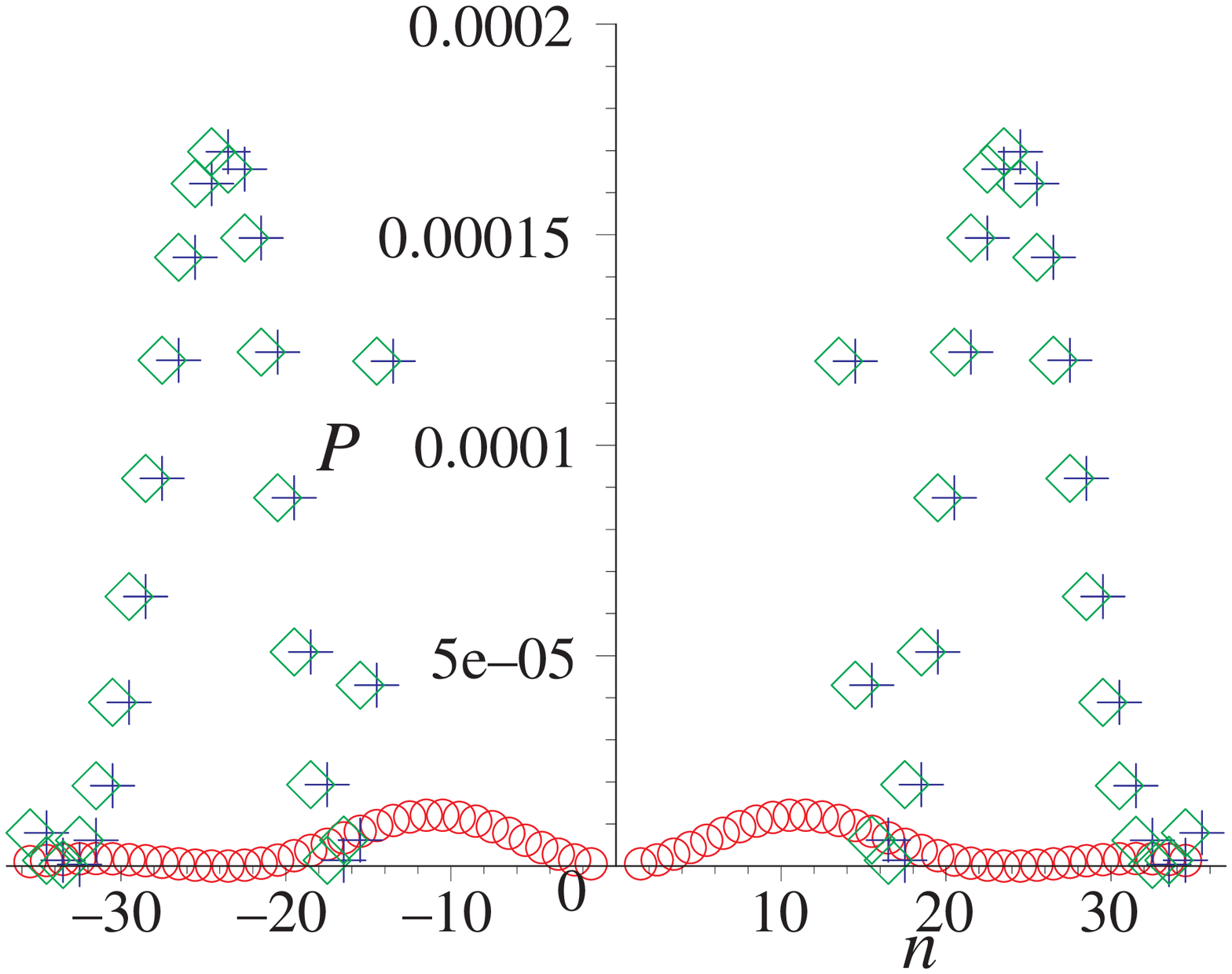}
\put(-220,0){\small \bf (a)} \put(-90,0){\small \bf (b)}
\caption{Transmission probabilities~$P=|t_n^2|$ as in
Fig.~\ref{figure4} above. When both slits are opened
simultaneously, red circles, the diffraction into higher orders is
very much suppressed {\bf (a)}. One has to zoom in, {\bf (b)}, to
see that the transmission probability $P=|t_n+t_{n+1}|^2$ (red
circles) is nonzero for orders higher than $\pm 1/2$.
\label{figure5}}
\end{figure}

Since amplitudes of adjacent orders have similar magnitudes but
alternating signs, see eq.~(\ref{r0rn}), the resulting
interference is destructive: opening the second slit in
stage~(iii) of the experiment counteracts the perturbation due to
the grating, which is so clearly visible in stage~(ii).
Figs.~\ref{figure4} and~\ref{figure5} illustrate the transition
from stage~(ii) (associated probability amplitudes $|r_n|^2$) to
stage~(iii) (associated amplitudes $|r_n+r_{n+1}|^2/2$). Afshar
erroneously interpreted this as a `complete' recovery of the slit
images to their pristine form encountered in stage~(i). Instead,
our analysis shows that in stage~(iii) light from the `wrong' slit
mostly compensates for the disturbances introduced by the grating
in stage~(ii) (although small residues of these perturbations
remain, see~\cite{Chown.04,Afshar05} and Figs.~\ref{figure4}~{\bf
(b)} and~\ref{figure5}~{\bf (b)}) but this is done at the expense
of redirecting light to the `wrong' detector.
%
%
%
%

Before closing this section let us revisit our assumptions. The
monochromaticity assumption is good for the narrow-band laser
light used in the experiment, the extension to temporal wave
packets is straightforward, so is an extension to wider slits. The
reduction to two dimensions applies to the plane depicted in
Fig.~\ref{fig1} and to a setup with long slits rather than round
holes, it does not affect the essence of the experiment. An
analysis of the exact setup described in Afshar's experiment is
only more tedious but not different in principle. Modelling the
grid by a planar reflecting film rather than round reflecting
wires allows us to write down expressions for transmitted and
reflected waves in simple form. This is obviously helpful for
tracking the flow of light in this setup and in a real experiment
this setup would perform better than the one used by Afshar
because none of the transmitted light gets reflected towards the
`wrong' detector it only reaches the wrong detector via
diffraction.

We can conclude that the orthodox interpretation of quantum
mechanics, described in this case with the help of classical wave
optics, is sufficient to analyze the experiment satisfactorily.
Our analysis is also applicable to some quantum cases such as
single photon, thermal or Glauber coherent light~\cite{Ole.PRA02}.
We will now quantify the complementarity behaviour in Afshar's
experiment.
%

\section{Complementarity in Afshar's experiment\label{sec.5.quantify.complementarity}}
The presence of the interference pattern in Afshar's experiment is
only inferred~\cite{Chown.04,Afshar05}, but according to the
orthodox interpretation of quantum mechanics it has to be measured
in order to be described by the quantum mechanical operator
formalism. This requirement is well captured by Wheeler's famous
dictum to the effect ``that only a registered event is a real
event"~\cite{Wheeler_dictum}. When studying the complementary
aspects of a quantum particle's behaviour both aspects have to be
measured simultaneously for every
particle~\cite{Wheeler.buch,Feynman.PhilosophicalImplications}.
How simultaneous `path' and `wave' measurements can be performed
with Afshar's setup is considered next.
%
\subsection{Qualitative discussion of complementarity}
%
The interference pattern at the position of the grid, in
stage~(iii), is responsible for the reduced scattering of photons
by the grid in stage~(iii) of the experiment. In order to measure
the contrast of the interference pattern without compromising the
simultaneous path detection we have to change the relative
phase~$\Delta \phi$ between light emanating from the two slits
(by, say, changing the relative path length between the two holes
with the help of a phase shifter, or, by moving the grating) and
then collect all light transmitted by the grating.

Note that transmitted and back-reflected photons form two mutually
exclusive subensembles to which the complementarity principle must
therefore be applied separately. In the following, just as in
Afshar's implementation of the experiment, we mostly deal with the
transmitted light only, but everything we say can analogously be
rephrased for the ensemble of back-reflected photons.

For transmitted light the finite slit widths of the grating reduce
the observed fringe contrast because the slits of the grating are
not infinitesimally small and therefore act as bucket detectors
rather than as the highly spatially resolving detectors required
for an ideal characterization of the interference pattern. This
leads to a tradeoff between path detection and determination of
the interference pattern contrast. Very narrow slits allow us to
observe the modulation due to the interference pattern with great
accuracy, but at the expense of strong diffraction of passing
photons, thus erasing their path information. For wider slits the
observed contrast of the interference pattern diminishes because
wide slits sample larger parts of the interference patter denying
good resolution, on the other hand the photon paths become less
disturbed yielding better path information.

For reflected photons the roles are reversed since small slits
correspond to wide reflecting stripes and vice versa.
%
\subsection{Quantification of Visibility}
%
The light in the grating plane forms a sinusoidal field
distribution, with the intensity distribution
$I(x)=\cos(\frac{\pi}{\Lambda}x+\Delta \phi)^2$, where $\Delta
\phi$ is the relative phase between the two slits~$S_1$ and~$S_2$.
To find out how much light gets transmitted we have to integrate
over the slit opening(s). We find that the transmitted intensity
is given by
$I_{t,max}=\int_{-\Lambda/2\cdot(1-a)}^{+\Lambda/2\cdot(1-a)} dx
\; \cos(\frac{\pi}{\Lambda}x)^2 = (\pi-a\pi+\sin(a \pi))/(2\pi)$
in the maximum case (grating positioned at interference pattern
minima) and
$I_{t,min}=\int_{-\Lambda/2\cdot(1-a)}^{+\Lambda/2\cdot(1-a)} dx
\; \sin(\frac{\pi}{\Lambda}x)^2 = (\pi-a\pi-\sin(a \pi))/(2\pi)$
in the minimum case (grating positioned at interference pattern
maxima). The ensuing measurable visibility of transmitted light
$V_t$ thus is
%
\begin{eqnarray}
V_t(a) \doteq \frac{I_{t,max} - I_{t,min}}{I_{t,max} + I_{t,min}}
= \frac{\sin(a \pi)}{\pi (1-a)} \; .
\label{Visibility}
\end{eqnarray}
%
An analogous calculation for reflected light is easily performed
and yields the expected result that the effective slit width in
this case is given by~$1-a$.
%
\subsection{Quantification of Distinguishability}

For a balanced interferometric setup such as Afshar's the
distinguishability~$\cal D$ of paths is determined by the power of
the path detectors $D_1$ and $D_2$ to discriminate the two paths.
Formally, it is given by half the distance between detector states
in the trace class norm~\cite{Englert96}. Describing the light
mode emanating from slit $S_j$ by the state $|S_j\rangle$ and the
mode associated with detector $D_k$ by state $|D_k\rangle$ we thus
have the following expression for the distinguishability of
transmitted light~${\cal D}_t$~\cite{Englert96}
%
\begin{eqnarray}
{\cal D}_t(a) & = & \frac{1}{2} \sum_{j=1}^2 \left| |\langle S_j |
D_1
\rangle|^2 - |\langle S_j | D_2 \rangle|^2 \right|  \\
& = & \frac{1}{2} \left( \left| |t_0|^2 - |t_1|^2 \right| +
\left| |t_1|^2 - |t_0|^2 \right| \right) \\
& = & (1 - a)^2  - \left( \frac{\sin(a\pi)}{\pi} \right)^2 \, .
\label{Distinguishibility}
\end{eqnarray}
%
\subsection{Quantification of Complementarity}
\begin{figure}[h t]
\centering
\includegraphics[width=1.8in,height=3.2in,angle=-90]{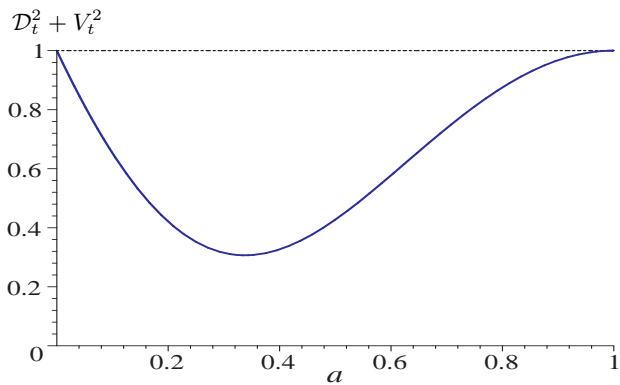}
\put(-230,5){\small ${\cal D}_t^2+V_t^2$}
\caption{Complementarity is never violated in Afshar's experiment:
${\cal D}_t^2(a) + V_t^2(a)\leq 1$ for all covering ratios $a$.
\label{figure6}}
\end{figure}

Combining these results allows us to see that
%
\begin{eqnarray}
{\cal D}_t^2(a) + V_t^2(a) \leq 1 \, ,
\label{Complementarity.checked}
\end{eqnarray}
%
where the limit of unity is only reached for the extreme cases of
$a=0$ (complete absence of a grid) or $a=1$ (complete covering by
a plane mirror that reflects all light and thus allows us to
discriminate the slits perfectly when utilizing the back-reflected
light), see Fig~\ref{figure6}.
%
%
\subsection{Some difficulties with Afshar's experiment}
%
Before we conclude let us mention some subtleties of the Afshar
experiment which could be quantified -- if one was so inclined.

The experiment is not very well designed to perform
complementarity experiments because the amount of transmitted and
reflected light depends on the relative position of the grating
with respect to the interference pattern. For energy conservation
reasons, the less light gets transmitted ($I_{t,min}$) the more
gets reflected and vice versa. This behaviour, in some sense
connects the subensemble of transmitted light with that of the
reflected light. Not in the sense that one subensemble allows us
to infer anything about the complementarity behaviour of the other
but in the sense that a more detailed analysis would have to
introduce weighting factors. This is an unwelcome complication I
neglected in my analysis.

Another weakness of the setup are losses, partly due to the use of
round wires in Afshar's implementation. But even in the simpler
version discussed here, we still encounter losses into higher
diffraction orders. Although they could be included into the
analysis they only diminish the path-resolution further and the
experimental setup would become more involved.

Finally, the greatest weakness in the analysis given by Afshar is
the \emph{inference} that an interference pattern must be present.
Quantum mechanics is not an ontological theory, it must not be
approached as such, only measured events are described by quantum
mechanics~\cite{Wheeler_dictum,Feynman.PhilosophicalImplications}.
%
\section{Conclusion}
The quantification of the principle of complementarity in
section~\ref{sec.5.quantify.complementarity} shows that Afshar's
experiment must not be interpreted as an example of a possible
violation of the principle of complementarity of quantum
mechanics. It is actually suboptimal in the sense that it does not
fully exhaust the limits stipulated by quantum mechanics; unlike,
say, the nearly optimal experiment performed by D\"{u}rr \emph{et
al.}~\cite{DuerrPRL98}.
\\

\begin{acknowledgments}
I wish to thank Dimitris Tsomokos for discussions.
\end{acknowledgments}
%


\begin{thebibliography}{25}
\bibitem[{pre()}]{press.echo}
\bibinfo{note}{Several more newspaper articles, weblogs, and online articles
  are devoted to dicussions of Afshar's experiment.}
\end{thebibliography}
\end{document}